\documentclass[9pt,shortpaper,twoside,web]{ieeecolor}
\usepackage{generic}

\usepackage{graphicx}
\usepackage{textcomp}

\usepackage{enumerate}
\usepackage{siunitx}
\usepackage{cite}

\usepackage[hidelinks]{hyperref}

\usepackage{amsmath, amssymb, mathrsfs, mathtools}

\newtheorem{thm}{Theorem}
\newtheorem{lem}{Lemma}
\newtheorem{prop}{Proposition}
\newtheorem{cor}{Corollary}
\newtheorem{defn}{Definition}
\newtheorem{prob}{Problem}

\newtheorem{assum}{Assumption}
\newtheorem{rem}{Remark}
\newtheorem{case}{Case}
\newtheorem{subcase}{Subcase}[case]

\def\BibTeX{{\rm B\kern-.05em{\sc i\kern-.025em b}\kern-.08em
    T\kern-.1667em\lower.7ex\hbox{E}\kern-.125emX}}
\begin{document}
\bstctlcite{MyBSTcontrol}

\title{Global Stabilization of Antipodal Points on $n$-Sphere with Application to Attitude Tracking}
\author{Xin Tong and Shing Shin Cheng
\thanks{Research reported in this work was supported in part by Innovation and Technology Commission of Hong Kong (ITS/136/20, ITS/135/20, ITS/234/21, ITS/233/21, and Multi-Scale Medical Robotics Center, InnoHK), in part by Research Grants Council (RGC) of Hong Kong (CUHK 24201219 and CUHK 14217822), in part by The Chinese University of Hong Kong (CUHK) Direct Grant 2021/2022, and in part by project BME-p7-20 of Shun Hing Institute of Advanced Engineering, CUHK. The content is solely the responsibility of the authors and does not necessarily represent the official views of the sponsors.}
\thanks{X. Tong is with Department of Mechanical and Automation Engineering, and CUHK T Stone Robotics Institute, The Chinese University of Hong Kong, Hong Kong. (e-mail: xtong@cuhk.edu.hk).}
\thanks{S. S. Cheng is with Department of Mechanical and Automation Engineering, CUHK T Stone Robotics Institute, Shun Hing Institute of Advanced Engineering, and Multi-Scale Medical Robotics Center, The Chinese University of Hong Kong, Hong Kong. (e-mail: sscheng@cuhk.edu.hk).}}

\maketitle

\begin{abstract}
    Existing approaches to robust global asymptotic stabilization of a pair of antipodal points on unit $n$-sphere $\mathbb{S}^n$ typically involve the non-centrally synergistic hybrid controllers for attitude tracking on unit quaternion space. However, when switching faults occur due to parameter errors, the non-centrally synergistic property can lead to the unwinding problem or in some cases, destabilize the desired set. In this work, a hybrid controller is first proposed based on a novel centrally synergistic family of potential functions on $\mathbb{S}^n$, which is generated from a basic potential function through angular warping. The synergistic parameter can be explicitly expressed if the warping angle has a positive lower bound at the undesired critical points of the family. Next, the proposed approach induces a new quaternion-based controller for global attitude tracking. It has three advantageous features over existing synergistic designs: 1) it is consistent, i.e., free from the ambiguity of unit quaternion representation; 2) it is switching-fault-tolerant, i.e., the desired closed-loop equilibria remain asymptotically stable even when the switching mechanism does not work; 3) it relaxes the assumption on the parameter of the basic potential function in literature. Comprehensive simulation confirms the high robustness of the proposed centrally synergistic approach compared with existing non-centrally synergistic approaches. 
\end{abstract}
  

\section{Introduction}

Stabilization of a pair of disconnected antipodal points on unit $n$-sphere $\mathbb{S}^n$ crops up in the field of robotics and aerospace applications. For instance, attitude control using quaternion representation exemplifies this task on $\mathbb{S}^3$ \cite{Verginis2020,Dong2022,Mason2022,Mattioni2022,Turner2021,Zhang2022}, since every rigid-body attitude on the special orthogonal group $\mathrm{SO(3)}$ corresponds to two antipodal quaternions on $\mathbb{S}^3$. In this work, we aim to design a new feedback controller to accomplish robust global stabilization of two antipodal points on $\mathbb{S}^n$ and thus achieve global attitude tracking on $\mathbb{S}^3$.

One major difficulty related to this control task, as shown in \cite{Sanfelice2006,Garone2010}, lies in that it is impossible to achieve robust global regulation to a set of disconnected points by using any continuous (even discontinuous) state feedback. For example, the sliding mode attitude maneuver controller in \cite{Dong2022} gives rise to a set of unwanted equilibria on the switching surface and trajectories starting from its neighborhood can yield slow convergence rate due to the vanishing state feedback. A common discontinuous strategy is to divide the sphere into two subspaces such that they are the basins of attraction of the two destination points, respectively \cite{Thienel2003,Mason2022}. However, arbitrarily small perturbations as formulated in \cite[Thm. 3.2]{Mayhew2011} can disorient the controller around the boundary of the subspaces.

Recently, the synergistic control in the framework of hybrid dynamical systems in \cite{Goebel2012} has emerged as a powerful tool to attain the robust global asymptotic stability \cite{Sanfelice2021}, which primarily relies on a synergistic family of potential functions. Roughly speaking, the synergistic property requires that for each function in the family and at each of its undesired critical point, there exists another potential function that has a lower value \cite{Mayhew2013,Berkane2017a}. The positive lower bound of the differences is called \emph{synergistic parameter}. Moreover, the family is called \emph{centrally} synergistic if all the potential functions are positive definite relative to the desired set; otherwise, it is \emph{non-centrally} synergistic. Then, the unwanted closed-loop equilibria can be avoided by hybrid switching mechanism: the state feedback associated to the minimum potential function is triggered when it leads to a decrease in the potential function that is greater than synergistic parameter.

The most commonly-used synergistic hybrid controller for stabilization of a pair of antipodal points on $\mathbb{S}^n$ consists of two continuous state feedback control laws, both of which stabilize one destination point while leaving the other destination point unstable \cite{Mayhew2011,Schlanbusch2012,Gui2016,Gui2018,Huang2021,Hashemi2021}. This design is non-centrally synergistic and thus suffers from switching fault. To be specific, if the switching does not work due to some malfunction, the desired set is guaranteed attractive but not Lyapunov stable, and thus may lead to the \emph{unwinding} phenomenon in attitude control \cite{Bhat2000}, namely yielding an unnecessary full rotation. Other synergistic control approaches on $\mathbb{S}^n$ in literature are designed for global regulation to only one setpoint such as \cite{Mayhew2010,Mayhew2013b,Casau2015a,Casau2019a}. Specifically, a collection of height functions that are positive definite relative to various setpoints was proposed with guaranteed non-centrally synergistic property \cite{Mayhew2013b}. The induced hybrid controller is non-robust to switching faults, because some of the control laws individually stabilize the plant to a setpoint far from the destination point. A centrally synergistic family was generated in \cite{Mayhew2010,Casau2015a} from a height function through \emph{angular warping}; however, the major technical problem is that the synergistic parameter essential for the synergistic control was not determined explicitly. In a nutshell, the existing synergistic approaches to setpoint regulation are not directly applicable to global stabilization of two antipodal points on $\mathbb{S}^n$.

It is worth mentioning that the existing approaches to attitude tracking face various additional challenges. First, it was shown in \cite{Bhat2000} that topological obstructions to global attitude tracking using continuous state feedback not only arise in the quaternion-based design but lie in the underlying state space $\mathrm{SO(3)}$ \cite{Raj2021,Akhtar2021,Invernizzi2020}. Second, the \emph{inconsistent} quaternion-based control laws, namely having different values for the quaternion representations of each rigid-body attitude, require a specific conversion mechanism to resolve the ambiguity of quaternion measurements \cite{Thienel2003,Mason2022,Mayhew2013a}. Otherwise, it may give rise to the troublesome unwinding (e.g., \cite{Verginis2020,Mattioni2022,Turner2021}) and chattering phenomenons. Finally, hybrid controllers have been developed on $\mathrm{SO(3)}$, with the non-centrally synergistic property in \cite{Mayhew2013,Berkane2017a} and centrally synergistic property in \cite{Berkane2017a,Berkane2017b,Wang2022}. Of note, the synergistic potential functions in \cite{Mayhew2013,Berkane2017a,Berkane2017b,Wang2022} were constructed from  the \emph{modified trace function} $P: \mathrm{SO(3)} \to \mathbb{R}$, given by $P(R) = \operatorname{tr} (\underline{A}(I-R))$ with constant symmetric matrix $\underline{A} \in \mathbb{R}^{3 \times 3}$. The major limitation of those approaches is that the conservative assumptions are imposed on the parameter: the matrix $\underline{A}$ is required to possess distinct eigenvalues in \cite{Mayhew2013} and single eigenvalue in \cite{Berkane2017b}; particularly, the matrix $\underline{A}$ is not allowed to have two largest eigenvalues---a typical scenario where there exists two noncollinear inertial vectors with equal weights for attitude measurement \cite{Berkane2017a}.

The main contributions of the present work are twofold. First, we develop the approach to generating the centrally synergistic potential functions on $\mathbb{S}^n$ relative to the antipodal points from a basic potential function via angular warping. To the best of our knowledge, it is the first centrally synergistic design for this task on $\mathbb{S}^n$ and also on unit quaternion space. Moreover, different from \cite{Mayhew2010,Casau2015a}, the proposed construction method is generic and offers explicit expression of the synergistic parameter based on the basic function parameter, which is desirable for the implementation of the hybrid controller. Second, we extend the centrally synergistic design to attitude tracking, yielding a new quaternion-based hybrid controller with three advantages: 1) it is consistent and thus can disambiguate the quaternion measurements that have to be handled carefully in the inconsistent design \cite{Verginis2020,Mason2022,Mattioni2022,Turner2021}; 2) it is switching-fault-tolerant in contrast to the non-centrally synergistic design in \cite{Mayhew2011,Schlanbusch2012,Gui2016,Gui2018,Huang2021,Hashemi2021,Mayhew2013b,Mayhew2013,Lee2015} and hence more robust in practice; 3) unlike the synergistic methods on $\mathrm{SO(3)}$ in \cite{Mayhew2013,Berkane2017a,Berkane2017b,Wang2022}, our approach is applicable to a modified potential function without extra requirement on its matrix parameter.

The rest of this article is organized as follows. Section~\ref{sec:pre} presents the preliminaries and the problem formulation. The main result is shown in Section~\ref{sec:main} and its application to attitude tracking is given in Section~\ref{sec:app}. Section \ref{sec:sim} presents some illustrative examples. Conclusions and perspectives are given in Section \ref{sec:con}. 

\section{Preliminaries and Problem Statement} \label{sec:pre}

\subsection{Notations and Lemmas}
We denote by $\mathbb{R}_{\geq 0}$ and $\mathbb{N}$, the sets of nonnegative real numbers and nonnegative integers, respectively. The standard Euclidean norm is defined as $|x| \coloneqq \sqrt{x^\top  x}$ for each $x \in \mathbb{R}^n$. The unit $n$-sphere is defined by $\mathbb{S}^n = \{ x \in \mathbb{R}^{n+1} : |x| = 1 \}$ and the \emph{tangent space} of $\mathbb{S}^n$ at $x \in \mathbb{S}^n$ is given by $\mathsf{T}_x \mathbb{S}^n = \{y \in \mathbb{R}^{n+1}: y^\top x = 0\}$. The $n$-dimensional closed ball with radius $r$ is $\bar{\mathbb{B}}_r^n = \{x \in \mathbb{R}^{n} : |x| \leq r\}$. For each symmetric matrix $A \in \mathbb{R}^{n \times n}$, we define $\mathcal{E} (A) = \{(\lambda,v) \in \mathbb{R} \times \mathbb{R}^n : A v = \lambda v, |v| = 1 \}$, $\mathcal{E}_\lambda (A) = \{\lambda \in \mathbb{R} : \exists (\lambda,v) \in \mathcal{E} (A) \}$, as well as $\mathcal{E}_v (A) = \{v \in \mathbb{R}^n : \exists (\lambda,v) \in \mathcal{E} (A) \}$, and denote by $\lambda_{\max}^A$ and $\lambda_{\min}^A$, the maximum and minimum of $\mathcal{E}_\lambda (A)$, respectively. Additionally, the geometric multiplicity of $\lambda \in \mathcal{E}_\lambda (A)$ is defined as $ \gamma_A (\lambda) = n - \mathrm{rank}(A - \lambda I)$. The kernel of $B \in \mathbb{R}^{m \times n}$ is given by $\ker (B) = \{x \in \mathbb{R}^n : B x = 0\}$. 
Given a finite set $\mathbb{Q} \subset \mathbb{N}$, we denote by $\mathcal{C}^1 (\mathbb{S}^n \times \mathbb{Q},\mathbb{R})$ the set of functions $U : \mathbb{S}^n \times \mathbb{Q} \to \mathbb{R}$ such that for each $q \in \mathbb{Q}$ the map $x \mapsto U(x,q) $ is continuously differentiable. Let $\nabla U(x,q) = [{\partial U(x,q)}/{\partial x^\top} ]^\top \in \mathbb{R}^{n+1}$ denotes the \emph{gradient} of $U$ with respect to the first argument. A function $U\in \mathcal{C}^1 (\mathbb{S}^n \times \mathbb{Q},\mathbb{R})$ is said to be \emph{positive definite} relative to a set $\mathcal{B} \subseteq \mathbb{S}^n \times \mathbb{Q}$ if $U(x,q) > 0$ for all $(x,q) \notin \mathcal{B}$, and $U(x,q) = 0$ if and only if $(x,q) \in \mathcal{B}$.

The state space of the rigid-body attitude is the \emph{special orthogonal group} of order three, $\mathrm{SO(3)} = \{R\in \mathbb{R}^{3\times 3}: R^\top R = I, \det R = 1\}$. Its Lie algebra is defined as $\mathfrak{so}(3) = \{X \in \mathbb{R}^{3\times 3}: X = - X^\top\}$. A rigid-body attitude can be represented by two antipodal points on $\mathbb{S}^3$, which is called unit quaternion and denoted by $Q = [\eta , \epsilon^\top]^\top \in \mathbb{S}^3$ with the scalar part $\eta \in \mathbb{R}$ and the vector part $\epsilon \in \mathbb{R}^3$. The identity quaternion is $\mathbf{i} = [1,0,0,0]^\top$. The rotation matrix is related to $Q$ through the mapping $\mathcal{R}_a: \mathbb{S}^3 \to \mathrm{SO}(3)$ defined by $\mathcal{R}_a(Q) = I + 2\eta \epsilon^\times + 2 ( \epsilon^\times)^2 $, where $(\cdot)^\times : \mathbb{R}^3 \to \mathfrak{so}(3)$ is from vector cross product such that $x^\times y = x \times y$, for all $x$, $y \in \mathbb{R}^3$. The quaternion multiplication is defined as $Q_1 \odot Q_2 = [\eta_1\eta_2 - \epsilon_1^\top \epsilon_2, \eta_1 \epsilon_2^\top + \eta_2 \epsilon_1^\top + (\epsilon_1 \times \epsilon_2)^\top]^\top$. Define the function $\nu : \mathbb{R}^3 \to \mathbb{R}^4$ by $\nu (x) = [0 , x^\top]^\top$.

The hybrid dynamical systems in \cite{Goebel2012,Sanfelice2021} are used. The notions of solutions to a hybrid system, hybrid time domain, and asymptotic stability are referred to \cite[Sec. 2.3.3 \& Sec. 3.2.1]{Sanfelice2021}. 
\begin{lem}[{\cite[Fact 4.14.7.]{Bernstein2018}}] \label{lem:pre_id}
    Let $n \geq 3 $ and $S \in \mathbb{R}^{n \times n}$ be skew-symmetric such that $S^3 = - a^2 S$ for some $a > 0$. Then, for all $\phi \in \mathbb{R}$, $e^{S \phi} \in \mathbb{R}^{n \times n}$ is an orthogonal matrix and $e^{S \phi} = I + a^{-1} \sin ( a \phi ) S + a^{-2} \bigl( 1 -  \cos ( a \phi ) \bigr) S^2$.
\end{lem}
\begin{lem}
Let $(e_1,\dots,e_n)$ be an orthonormal basis of $\mathbb{R}^n$, and $\mathcal{V} = \bigcup_{2 \leq i \leq n} \{e_i, -e_i\}$. Then, the following inequalities hold.
{\small
\begin{subequations}
    \begin{align}
        n \max_{i} (e_i^\top x)^2 & \geq x^\top x , \ \forall n \geq 1, \label{eq:useful_ineq2} \\
        \max_{ e \in \mathcal{V} } \left| x^\top (e + c e_1) \right| & \geq \sqrt{ \frac{x^\top x- (x^\top e_1)^2}{n-1} } + c \left|x^\top e_1 \right| , \forall n \geq 2 \label{eq:useful_ineq1}
    \end{align}
\end{subequations}}%
for all $x \in \mathbb{R}^n $ and $c \geq 0$.
\end{lem}
\begin{proof}
    Write $x = \sum_{i=1}^{n} a_i e_i$ for $a_i \in \mathbb{R}$. The inequalities \eqref{eq:useful_ineq2} and \eqref{eq:useful_ineq1} are obvious from the inequalities $n \max_i a_i^2 \geq \sum_{i=1}^n a_i^2$ and $\max_{ 2 \leq i \leq n} |a_i|  \geq (\frac{1}{n-1}\sum_{i=2}^n a_i^2)^{1/2} $, respectively.
\end{proof}

\subsection{Problem Formulation}
The dynamics of the system evolving on $\mathbb{S}^n$ can be represented by 
\begin{align}
    \dot{x} & = \Pi (x) \omega,  & x \in \mathbb{S}^n, \label{eq:dyn_sphere}
\end{align}
where $\omega \in \mathbb{R}^{n+1}$ is the input and $\Pi:\mathbb{S}^n \to \mathbb{R}^{(n+1) \times (n+1)}$ defined by $\Pi (x) = I  - x x^\top$ maps $\omega$ onto $ \mathsf{T}_x \mathbb{S}^n$. 

Let $\mathbb{Q} \subset \mathbb{N}$ be a nonempty set. We use the function $U \in \mathcal{C}^1 (\mathbb{S}^n \times \mathbb{Q},\mathbb{R})$ to encapsulate the family of potential functions that is indexed by the logical variable $q \in \mathbb{Q}$. The set of critical points of $U$ is given by $\operatorname{Crit} {U} = \{ (x,q) \in \mathbb{S}^n \times \mathbb{Q} : \Pi (x) \nabla U(x,q) = 0\}$.
\begin{defn}[{\cite{Mayhew2013,Mayhew2013b}}] \label{def:cenSynSphere}
    Let $\mathbb{Q} \subset \mathbb{N}$ be a nonempty finite set and $r \in \mathbb{S}^n$ be the reference point. Define the sets
    \begin{align}
        \mathcal{A}_0 &= \{ r,-r\}, & \mathcal{B}_0 &= \mathcal{A}_0 \times \mathbb{Q}. \label{eq:setpoint_def}
    \end{align}
    Let $U \in \mathcal{C}^1 (\mathbb{S}^n \times \mathbb{Q},\mathbb{R})$ be positive definite relative to a nonempty subset $\mathcal{B}_1 \subseteq \mathcal{B}_0$. The \emph{synergy gap} of $U$ is defined by the function $\mu_U \in \mathcal{C}^1 (\mathbb{S}^n \times \mathbb{Q},\mathbb{R})$ such that $\mu_U(x,q) = U(x,q) - \min_{p\in \mathbb{Q}} U(x,p)$. Then, $U $ is called \emph{synergistic} relative to $\mathcal{A}_0$ if there exist two functions $\delta : \mathbb{Q} \to \mathbb{R}_{\geq 0}$ and $\bar{\delta} : \mathbb{Q} \to \mathbb{R}_{\geq 0}$ such that
    \begin{align} \label{eq:cenSyn_def}
        \forall (x,q) \in  (\operatorname{Crit} {U} \bigcup \mathcal{B}_0 )\setminus \mathcal{B}_1, \; \mu_U(x,q) & \geq \bar{\delta} (q) > \delta (q) > 0,
    \end{align}
    in which case, $\delta$ is called \emph{synergistic parameter} and $U$ is called synergistic with \emph{gap exceeding} $\delta$. In addition, if $\mathcal{B}_1 = \mathcal{B}_0$ ($\mathcal{B}_1 \subsetneq \mathcal{B}_0$), $U $ is called \emph{centrally (non-centrally)} synergistic relative to $\mathcal{A}_0$.
\end{defn}
\begin{rem}
    A synergistic hybrid controller consists of the gradient-descent feedbacks induced from the synergistic potential functions and by the synergistic property \eqref{eq:cenSyn_def}, guarantees that the system \eqref{eq:dyn_sphere} can be pushed away from the unwanted critical points and be asymptotically stabilized on $\mathcal{B}_1$ by switching to the state feedback associated to the minimum potential function in the family. 
\end{rem}
\begin{prob} \label{prob:1}
    Construct the centrally synergistic potential functions relative to the set $\mathcal{A}_0$ of \eqref{eq:setpoint_def} and thereafter design a hybrid controller to robustly globally asymptotically stabilize the system of \eqref{eq:dyn_sphere} to the set $\mathcal{A}_0$.
\end{prob}

A natural extension of Problem~\ref{prob:1} is quaternion-based global attitude tracking, since the unit quaternion space double cover $\mathrm{SO(3)}$. Consider a rigid body system described by 
\begin{align} 
    \begin{cases}
        \dot{Q}  = \frac{1}{2} Q \odot \nu (\omega) = \frac{1}{2} \Lambda (Q) \omega, \\
    J\dot{\omega} = - \omega^\times J \omega + \tau,
    \end{cases} \label{eq:body_traj}
\end{align}
where the unit quaternion $Q \in \mathbb{S}^3$ is the rigid-body attitude, $\omega \in \mathbb{R}^3$ is the body-frame angular velocity, $J = J^\top \in \mathbb{R}^{3\times 3}$ is the inertia matrix, $\tau \in \mathbb{R}^3$ is an external torque, and the function $\Lambda : \mathbb{S}^3 \to \mathbb{R}^{4 \times 3}$ defined by $\Lambda (Q) = [-\epsilon, \eta I - \epsilon^\times]^\top$ projects $\omega$ onto $\mathsf{T}_Q \mathbb{S}^3$. 

Let $c_\omega, c_a > 0$ be constant, and consequently, $\mathcal{W}_d \coloneqq \mathbb{S}^3 \times \bar{\mathbb{B}}^3_{c_\omega}$ is compact. The reference trajectory is generated by the following dynamical system \cite{Mayhew2011,Mayhew2013,Wang2022}
\begin{equation} \label{eq:ref_traj}
    \begin{rcases}
        \dot{Q}_d = \frac{1}{2} Q_d \odot \nu (\omega_d ) \\
        \dot{\omega}_d \in \bar{\mathbb{B}}_{c_a}^3
    \end{rcases} 
    (Q_d,\omega_d) \in \mathcal{W}_d .
\end{equation}
The error quaternion and error velocity can then be defined as $\tilde{Q}= Q_d^{-1} \odot Q$ and $\tilde{\omega} = \omega - \mathcal{R}_a (\tilde{Q})^\top \omega_d$. In addition, let $\bar{\omega}_d \coloneqq \mathcal{R}_a (\tilde{Q})^\top \omega_d $. 
Combining \eqref{eq:body_traj} and \eqref{eq:ref_traj} yields the error dynamics
\begin{equation}
    \begin{cases}
        \dot{\tilde{Q}} = \frac{1}{2} \tilde{Q} \odot \nu (\tilde{\omega}), \\
        J \dot{\tilde{\omega}} = \Sigma (\tilde{Q},\tilde{\omega},\bar{\omega}_d) \tilde{\omega} - \Xi (\tilde{Q},\bar{\omega}_d,\dot{\omega}_d) + \tau,
    \end{cases} \label{eq:right_err_traj}
\end{equation}
where the functions $\Sigma :  \mathbb{R}^3 \times \mathbb{R}^3 \to \mathfrak{so}(3)$ and $\Xi : \mathbb{S}^3 \times \mathbb{R}^3 \times \mathbb{R}^3 \to \mathbb{R}^3 $ are given by $\Sigma(\tilde{Q},\tilde{\omega},\bar{\omega}_d) = \bigl(J (\tilde{\omega}+ \bar{\omega}_d) \bigr)^\times - \bar{\omega}_d^\times J  - J \bar{\omega}_d^\times $ and $\Xi (\tilde{Q},\bar{\omega}_d,\dot{\omega}_d) = J \mathcal{R}_a (\tilde{Q})^\top \dot{\omega}_d + \bar{\omega}_d^\times J \bar{\omega}_d$, respectively.

Define the state space $\mathcal{W}_z \coloneqq \mathcal{W}_d \times \mathbb{S}^3 \times \mathbb{R}^3 $ and the state $ z \coloneqq (Q_d, \omega_d, \tilde{Q}, \tilde{\omega}) \in \mathcal{W}_z$. Then, we describe the problem of global attitude tracking as follows.
\begin{prob} \label{prob:2}
    Design a controller such that for all initial conditions $z \in \mathcal{W}_z$, trajectories $z(t)$ asymptotically approach the set $ \mathcal{A}_1 \coloneqq \{z \in \mathcal{W}_z : \tilde{Q} \in \{\mathbf{i},-\mathbf{i} \}, \tilde{\omega} = 0 \}$ for the closed-loop system.
\end{prob}

The following hybrid controller proposed in \cite{Mayhew2011} may be the most popular quaternion solution to Problem~\ref{prob:2} 
\begin{subequations}
    \begin{align}
        & \tau  = \Xi (\tilde{Q},\bar{\omega}_d,\dot{\omega}_d)   -  k_1 q \tilde{\epsilon} - k_2  \tilde{\omega} , \label{eq:NonCS_ctr_tau} \\
        & 
        \begin{cases}
        \dot{q} = 0 & (z,q) \in \{ (z,q) \in \mathcal{W}_z \times \mathbb{Q} :  q \tilde{\eta} \geq - \delta \}   ,\\
        q^+ = - q & (z,q) \in \{ (z,q) \in \mathcal{W}_z \times \mathbb{Q} :  q \tilde{\eta} \leq - \delta \} , \label{eq:NonCS_ctr_q}
        \end{cases}
    \end{align} 
    \label{eq:NonCS_ctr_quat}
\end{subequations}
where $\mathbb{Q} = \{-1,1\}$ and $k_1,k_2 > 0$. However, \eqref{eq:NonCS_ctr_quat} is a non-centrally synergistic design in the manner that the control law for each $q \in \mathbb{Q}$ cannot stabilize $\mathcal{A}_1$ individually. Consequently, the unwinding can arise with switching faults. Of note, \eqref{eq:NonCS_ctr_quat} led to its variants in \cite{Schlanbusch2012,Gui2016,Gui2018,Huang2021,Hashemi2021}. This motivates our centrally synergistic design for Problem~\ref{prob:2}.

\section{Main Results} \label{sec:main}

This section first takes a gradient-descent feedback that is generated from a basic potential function as an example to show the challenges of using continuous feedback control to stabilize the system \eqref{eq:dyn_sphere}. Then, we demonstrate that the synergistic hybrid controller derived from a generic centrally synergistic family of potential functions can help to achieve global stabilization. Finally, we propose a systematic approach to constructing a centrally synergistic family from the basic potential function, so as to realize the synergistic control.

\subsection{Continuous Feedback Control}
Consider the function $P : \mathbb{S}^n \to \mathbb{R}$ defined by 
\begin{equation}
    P (x) = x^\top M x, \label{eq:PF_def}
\end{equation}
where $M \in \mathbb{R}^{(n+1) \times (n+1)}$ is symmetric and positive semidefinite. Since the eigenvalues of a symmetric matrix are real, henceforth we adopt the convention that the eigenvalues of $M$ are always arranged in a nondecreasing order. The next assumption guarantees that $P$ is a basic potential function relative to $\mathcal{A}_0$. We call it basic since it will be used to construct the synergistic potential functions.
\begin{assum} \label{ass:M_cond}
    The symmetric matrix $M$ in \eqref{eq:PF_def} has the eigenvalues as $\lambda_{\min}^M = \lambda_0 < \lambda_1 \leq \lambda_2 \leq \dots \leq \lambda_n = \lambda_{\max}^M $ and $\lambda_0 = 0$ is associated with the unit eigenvector $v_0 = r$.
\end{assum}

\begin{lem}\label{lem:crit_P}
    Consider the function $P$ given by \eqref{eq:PF_def}. Its set of critical points is given by $\operatorname{Crit} {P} = \mathcal{E}_v (M)$. 
\end{lem}
\begin{proof}
    The gradient of $P$ is given by $\nabla P (x) = 2Mx$. It follows that $\operatorname{Crit} {P} = \{x\in \mathbb{S}^n: 2\Pi (x) M x = 0 \}$. Note that $\Pi(x) y = 0 $ for $x \in \mathbb{S}^n$ and $y \in \mathbb{R}^{n+1}$ if and only if $x$ and $y$ are collinear or $y = 0$. Hence, $\operatorname{Crit} {P} = \mathcal{E}_v (M) \bigcup ( \ker (M) \bigcap \mathbb{S}^n )$. If $M$ has zero eigenvalue, $\ker (M)$ is the eigenspace of $M$ associated with $\lambda = 0$; otherwise, $\ker (M) = \emptyset$. In consequence, $\operatorname{Crit} {P} = \mathcal{E}_v(M)$.
\end{proof}

The lemma implies that global convergence cannot be achieved by the continuous gradient-descent feedback from the potential function \eqref{eq:PF_def}, because the feedback vanishes at its inevitable, undesired critical points. Moreover, such continuous (or even pure discontinuous) feedback may fail to achieve stabilization to $\mathcal{A}_0$ in the presence of oscillating noise that changes the controller's target on which way to stabilize \cite{Sanfelice2006,Garone2010}. The next proposition describes such properties.

\begin{prop} \label{prop:AGAS_clp}
    Let Assumption~\ref{ass:M_cond} hold. Define the function $\kappa_0 : \mathbb{R}^{n+1} \to \mathbb{R}^{n+1}$ as $\kappa_0 (x) = - 2 c_0 \Pi(x) M x$ with $c_0 > 0$.
    \begin{enumerate}
        \item For the closed-loop system consisting of the control law $\omega = \kappa_0(x)$ and the system \eqref{eq:dyn_sphere}, the set $(\mathcal{E}_v (M) \setminus \mathcal{A}_0)$ is forward invariant and $\mathcal{A}_0$ is locally asymptotically stable. 
        \item Let $\alpha > 0$ and $\mathcal{A}_0^c = \bigl((\mathcal{E}_v (M) \setminus \mathcal{A}_0) + \bar{\mathbb{B}}_{\alpha}^{n+1} \bigr) \bigcap \mathbb{S}^n$. For each initial conditions $x(0) \in \mathcal{A}_0^c $, there exist a piecewise constant function $n_{\alpha} : [0,\infty) \to \bar{\mathbb{B}}_{\alpha}^{n+1}$ and a Caratheodory solution $x $ to the closed-loop system consisting of $\omega = \kappa_0(x+n_{\alpha})$ and \eqref{eq:dyn_sphere} satisfying $x(t) \in \mathcal{A}_0^c$ for all $t \in [0,\infty)$.
    \end{enumerate}
\end{prop}
\begin{proof}
    In the item 1), the closed-loop system is governed by $\dot{x} = - 2c_0 \Pi(x) M x$, where we have used the identity $\Pi(x) \Pi (x) = \Pi (x)$ for all $x \in \mathbb{S}^n$. It is obvious that the set $(\mathcal{E}_v (M) \setminus \mathcal{A}_0)$ is forward invariant. The stability of $\mathcal{A}_0$ can be obtained by using \eqref{eq:PF_def} as Lyapunov function. The item 2) follows from \cite[Thm. 2.6]{Sanfelice2006}.
\end{proof}

\subsection{Synergistic Control}
We shall showcase how the synergistic control can be used to address the challenges encountered by continuous feedback control.

Given a function $U \in \mathcal{C}^1 (\mathbb{S}^n \times \mathbb{Q},\mathbb{R})$ centrally synergistic relative to $\mathcal{A}_0$, we consider the state feedback with some $c_1 > 0$
\begin{equation}
    \kappa_1 (x,q) = - c_1 \Pi(x) \nabla U(x,q) . \label{eq:hybrid_ctr}
\end{equation}
Applying the control law \eqref{eq:hybrid_ctr} to the system \eqref{eq:dyn_sphere} results in the hybrid closed-loop as 
\begin{equation} \label{eq:hybrid_clp1}
    \mathcal{H}_{1} : \ 
    \begin{cases}
        \begin{rcases}
            \dot{x} = \Pi (x) \kappa_1 (x,q) \\ 
            \dot{q} = 0
        \end{rcases} 
        & (x,q) \in \mathcal{F}_{1}  ,\\
        \begin{rcases}
            x^+ = x  \\
            q^+ \in G_{1} (x,q)
        \end{rcases} 
        & (x,q) \in \mathcal{J}_{1} ,
    \end{cases}
\end{equation}
where the jump map $G_1: \mathbb{S}^n \times \mathbb{Q} \rightrightarrows \mathbb{Q}$ is defined by $G_1 (x,q) = \arg\min_{p \in \mathbb{Q}} U(x,p)  $, the flow and jump sets are given by $\mathcal{F}_{1} = \{ (x,q) \in \mathbb{S}^n \times \mathbb{Q}  : \mu_U (x,q)\leq \delta(q) \}$ and $\mathcal{J}_{1}  = \{ (x,q) \in \mathbb{S}^n \times \mathbb{Q}  : \mu_U (x,q)\geq \delta(q)  \}$, respectively. Note that the hybrid closed-loop system \eqref{eq:hybrid_clp1} is autonomous and satisfies the hybrid basic conditions \cite[Assumption 6.5]{Goebel2012}. The next proposition states that Problem~\ref{prob:1} can be solved by \eqref{eq:hybrid_clp1}.

\begin{prop} \label{prop:gas_sphere}
    Let $U \in \mathcal{C}^1 (\mathbb{S}^n \times \mathbb{Q},\mathbb{R})$ be centrally synergistic relative to $\mathcal{A}_0 $ of \eqref{eq:setpoint_def}. Then, the following statements hold.
    \begin{enumerate}[1)]
        \item $\mathcal{B}_0$ is globally asymptotically stable for $\mathcal{H}_1$ of \eqref{eq:hybrid_clp1}.
        \item $\mathcal{B}_0$ is semiglobally practically robustly $\mathcal{KL}$ asymptotically stable for the nominal system $\mathcal{H}_1$ of \eqref{eq:hybrid_clp1}. Especially, there exists a class-$\mathcal{KL} $ function $\beta$, such that for each $\varepsilon > 0$ and each compact set $\mathcal{K} \subset \mathbb{S}^n \times \mathbb{Q}$ there exists $\rho > 0$ such that each solution $(x,q)$ to the closed-loop system with measurable disturbance $n_d : [0,\infty) \to {\bar{\mathbb{B}}}^{n+1}_\rho$ from $\mathcal{K}$ governed by 
        \begin{equation*} 
            \mathcal{H}_{1}^* : \ 
            \begin{cases}
                \begin{rcases}
                    \dot{x} = \Pi (x) \kappa_1 (x+n_d,q) \\ 
                    \dot{q} = 0
                \end{rcases} 
                & (x+n_d,q) \in \mathcal{F}_{1}  ,\\
                \begin{rcases}
                    x^+ = x  \\
                    q^+ \in G_{1} (x + n_d,q)
                \end{rcases} 
                & (x + n_d,q) \in \mathcal{J}_{1} ,
            \end{cases}
        \end{equation*}
        satisfies that $|x(t,j)|_{\mathcal{A}_0}  \leq \beta \bigl(|x(0,0)|_{\mathcal{A}_0} , t+j  \bigr) + \varepsilon $ holds for all $(t,j) \in \operatorname{dom} x$.
    \end{enumerate}
\end{prop}
\begin{proof}
    Invoking the hybrid Lyapunov theorem \cite[Thm. 3.19]{Sanfelice2021}, the item (1) is easily shown by using $U$ as a Lyapunov function candidate. The item (2) is established by \cite[Lemma 7.20]{Goebel2012}.
\end{proof}
\begin{rem}
    Proposition \ref{prop:gas_sphere} states that the synergistic hybrid controller in the presence of $\rho$-size disturbances can regulate the state $x$ to $\varepsilon$ close to $\mathcal{A}_0$ from arbitrary set of initial conditions $\mathcal{K}$. 
\end{rem}

\subsection{Construction of a Centrally Synergistic Family on \texorpdfstring{$\mathbb{S}^n$}{n-Sphere}}  
We shall show how to construct the centrally synergistic potential functions upon the basic function \eqref{eq:PF_def}, so as to implement the synergistic hybrid controller. The next lemma introduces a useful tool called the \emph{angular warping} and its proof is deferred to Appendix \ref{pf:nec_cond_synFunc}.

\begin{lem} \label{lem:nec_cond_synFunc}
    Let $\mathbb{Q} \subset \mathbb{N}$ be a nonempty finite subset and assign a skew-symmetric matrix $S_q \in \mathbb{R}^{(n+1)\times (n+1)} $ to each $q \in \mathbb{Q}$. Let $\theta: \mathbb{S}^n \to \mathbb{R}_{\geq 0} $ be a real-valued differentiable function. Consider the \emph{warping function} $\mathcal{T} : \mathbb{S}^n \times \mathbb{Q} \to \mathbb{S}^n$ defined by 
    \begin{align}
        \mathcal{T} (x,q) = e^{S_q \theta (x)} x. \label{eq:warp_def}
    \end{align}
    Then, the gradient of $\mathcal{T}$ is given by\footnote{We define the gradient of $\mathcal{T}$ as the transpose of the Jacobian matrix of $\mathcal{T}$ with respect to its first argument.}
    \begin{align}
        \nabla \mathcal{T}(x,q) &= \bigl( I - \nabla \theta (x) x^\top S_q  \bigr) e^{-S_q \theta(x)}. \label{eq:grad_warp} 
    \end{align}
    If $\det \bigl(\nabla \mathcal{T}(x,q) \bigr) \neq 0$ for all $(x,q) \in \mathbb{S}^n \times \mathbb{Q}$, then the mapping $x \mapsto \mathcal{T}(x,q)$ is everywhere a local \emph{diffeomorphism}. Additionally, if $V : \mathbb{S}^n  \to \mathbb{R} $ is differentiable and positive definite relative to the set $\mathcal{A}_0$ and $\mathcal{T}^{-1} (\mathcal{A}_0) = \mathcal{B}_0$, then the composite function $U = V \circ \mathcal{T} \in \mathcal{C}^1 (\mathbb{S}^n \times \mathbb{Q},\mathbb{R})$ is positive definite relative to $\mathcal{B}_0$, and the set of critical points of $U$ is given by $\operatorname{Crit} {U} = \mathcal{T}^{-1}(\operatorname{Crit} {V})$.
\end{lem}

\begin{rem}
    Lemma~\ref{lem:nec_cond_synFunc} shows that given a potential function $V$, a new potential function $x \mapsto U(x,q)$ can be generated by compositing a diffeomorphic warping transformation $\mathcal{T}$ such that its critical points are different from $V$. Therefore, it is natural to consider a family of potential functions generated through various warping directions as a candidate satisfying the centrally synergistic property, which necessarily requires $\mathcal{T}^{-1} (\mathcal{A}_0) = \mathcal{B}_0$ by this lemma. Finally, Lemma~\ref{lem:nec_cond_synFunc} takes effect for a generic set $\mathcal{A}_0$ and thus provide a more general construction tool on $\mathbb{S}^n$ than \cite{Mayhew2010,Casau2015a}. 
\end{rem}

We now define some parameters for the construction of the synergistic potential functions. Let Assumption \ref{ass:M_cond} hold, $\mathbb{Q} = \{1,\dots, 2n\}$, and the unit eigenvectors $(v_1, \dots, v_n)$ associate to the eigenvalues $(\lambda_1, \dots, \lambda_n)$ of $M$. Then, we define the extended eigenvalues $\lambda_q$, the vector $u_q \in \mathbb{S}^n$, and the skew-symmetric matrix $S_q \in \mathbb{R}^{(n+1)\times (n+1)}$ for $q \in \mathbb{Q}$ as 
\begin{align} 
    \lambda_{q+n} &= \lambda_{q}, &  q &\in \{1,\dots,n\}, \\
    u_q &= - u_{q+n} = v_q, &  q &\in \{1,\dots,n\}, \label{eq:u_q_def} \\
    S_q &= u_q r^\top - r u_q^\top,  & q &\in \mathbb{Q} ,\label{eq:Sq_def}
\end{align}
where the definition of $\lambda_q$ is extended to correspond to the eigenvector $u_q$ with $ q > n$. Define the index bijective function $\mathfrak{q}: \mathbb{Q} \to \mathbb{Q}$ by $\mathfrak{q} (q) = \{q-n, q+n\} \bigcap \mathbb{Q}$. It follows that $u_q = -u_{\mathfrak{q}(q)}$ for $q \in \mathbb{Q}$. Define the index set corresponding to the eigenvalue $\lambda$ by $\mathbb{Q}_\lambda = \{q \in \mathbb{Q}: \lambda_q = \lambda \}$. By definition of \eqref{eq:u_q_def}, the set $\bigcup_{p \in \mathbb{Q}_{\lambda}} \{u_p\}$ is composed of an orthonormal basis of the eigenspace $M$ associated with $\lambda$ and the negative of the basis.

The next theorem states the approach to constructing a centrally synergistic family from the basic potential function by using a certain warping angle function. The proof is deferred to Appendix \ref{pf:syn_gap_positive}.

\begin{thm} \label{thm:syn_gap_positive}
    Let $\mathbb{Q} = \{1,\dots, 2n\}$. Consider the functions $P$ of \eqref{eq:PF_def} under Assumption~\ref{ass:M_cond}, and $\mathcal{T}$ of \eqref{eq:warp_def} with the skew-symmetric matrices given by \eqref{eq:Sq_def} and the function $\theta$, such that $\theta$ is positive definite relative to $\mathcal{A}_0$ and $\theta (x) < \frac{\pi}{4}$ for all $x \in \mathbb{S}^n$. Suppose that $\det \bigl( \nabla \mathcal{T}(x,q) \bigr) \neq 0$ for all $(x,q) \in \mathbb{S}^n \times \mathbb{Q}$ and $\mathcal{T}^{-1}\bigl( \mathcal{A}_0 \bigr) = \mathcal{B}_0$. If in addition there exists $\underline{\vartheta} > 0$ such that $\theta (x) \geq \underline{\vartheta}$ for all $(x,q) \in \operatorname{Crit} {U}\setminus \mathcal{B}_0$, then the composite function $U = P \circ \mathcal{T} $ is centrally synergistic relative to $\mathcal{A}_0$ and the synergistic parameter $\bar{\delta}$ is given by 
    \begin{align}
        \bar{\delta}(q) = \min \bigl\{ \Delta_1(q) , \Delta_2(q) \bigr\} \label{eq:bar_delta_def}
    \end{align}
    where $\Delta_1: \mathbb{Q} \to \mathbb{R}_{\geq 0}$ and $\Delta_2 : \mathbb{Q} \to \mathbb{R}_{\geq 0}$ are defined by
    \begin{align}
        \Delta_1 (q) &= \min_{\lambda \in \mathcal{E}_\lambda (M) \setminus \{\lambda_q,0 \} }  \frac{\sin^2 (\underline{\vartheta})}{\gamma_M (\lambda)} \lambda, \label{eq:bar_delta_def_1} \\
        \Delta_2 (q) &= 
        \begin{cases}
            \lambda_q \sin^2 (2 \underline{\vartheta}) , & \gamma_M(\lambda_q) = 1, \\
            \lambda_q \min \Bigl\{ \frac{\sin^2 ( \underline{\vartheta} )}{\gamma_M(\lambda_q) - 1}    ,\frac{\sin^2 ( 2\underline{\vartheta})}{4}    \Bigr\}, & \gamma_M(\lambda_q) > 1.
        \end{cases}
        \label{eq:bar_delta_def_2}
    \end{align}
\end{thm}
\begin{rem}
    By Theorem \ref{thm:syn_gap_positive}, if the angular angle function $\theta$ has a positive lower bound at $\operatorname{Crit} {U} \setminus \mathcal{B}_0$, the synergistic condition \eqref{eq:cenSyn_def} is guaranteed and $\bar{\delta}$ is explicitly specified by \eqref{eq:bar_delta_def}. In addition, since $S_q$ defined by \eqref{eq:Sq_def} satisfies $S_q^3 = -S_q$, it follows from Lemma \ref{lem:pre_id} 
    \begin{equation}
        e^{S_q \phi}  = I + \sin (\phi) S_q  + \bigl(1 - \cos (\phi)\bigr) S_q^2   \label{eq:id3}
    \end{equation}
    which is more suitable for implementing \eqref{eq:warp_def}.
\end{rem}

Next, we show that the basic potential function of \eqref{eq:PF_def} can be used as the warping angle candidate. The proof is given in Appendix \ref{pf:syn_gap_sol}.

\begin{thm} \label{thm:syn_gap_sol}
    Continuing with all assumptions and variables defined as in Theorem \ref{thm:syn_gap_positive}, we define the warping angle function $\theta: \mathbb{S}^n \to \mathbb{R}_{\geq 0}$ as $\theta (x) =  k \lambda_n^{-1} P(x) $ from \eqref{eq:PF_def}, where $ 0 < k < \frac{\pi}{4} $ is a constant gain. Define the function $\Theta : \mathbb{Q} \to \mathbb{R}_{\geq 0}$ as  
\begin{equation}
    \Theta(q) = \frac{2 k \lambda_q}{\lambda_n + \sqrt{\lambda_n^2 + 4k^2 \lambda_q^2}}.
\end{equation}
Then, the following statements hold.
\begin{enumerate}
    \item $\det \bigl( \nabla \mathcal{T}(x,q)  \bigr) \neq 0$ for all $(x,q) \in \mathbb{S}^n \times \mathbb{Q}$ and $\mathcal{T}^{-1} \bigl( \mathcal{A}_0 \bigr) = \mathcal{B}_0$.
    \item The composite function $U = P \circ \mathcal{T}$ is centrally synergistic relative to $\mathcal{A}_0$, and $\bar{\delta}$ defined in \eqref{eq:cenSyn_def} is given by
    \begin{subequations} \label{eq:bar_delta_sol}
    \begin{align}
        &\bar{\delta}(q) = \min \bigl\{ \Delta_1(q) , \Delta_2(q) \bigr\} ,\\
        &\Delta_1 (q) = \min_{\lambda \in \mathcal{E}_\lambda (M) \setminus \{\lambda_q,0 \} }  \frac{\sin^2 (k \lambda / \lambda_n)}{\gamma_M (\lambda)} \lambda  , \\
        &\Delta_2 (q) = 
        \begin{cases}
            \lambda_q \sin^2 (2 \Theta(q)) , & \gamma_M(\lambda_q) = 1, \\
            \lambda_q \min \Bigl\{ \frac{\sin^2 ( \Theta(q) )}{\gamma_M(\lambda_q) - 1}    ,\frac{\sin^2 ( 2 \Theta(q))}{4}    \Bigr\}, & \gamma_M(\lambda_q) > 1.
        \end{cases} 
    \end{align}
    \end{subequations}
\end{enumerate}
\end{thm}
\begin{rem}
    The gain $k$ is chosen to guarantee the warping transformation as ``good'' as described in Lemma \ref{lem:nec_cond_synFunc}. The explicit expression of the upper bound in \eqref{eq:bar_delta_sol} makes the hybrid controller in \eqref{eq:hybrid_clp1} easier to implement. The analogous tools developed for the desired set consisting of a single point in \cite{Mayhew2010,Casau2015a} do not uncover the computation of such critical technical parameter.
\end{rem}

\section{Application to Attitude Tracking} \label{sec:app}
This section shows how to apply the theoretical results to global attitude tracking. First, we detail the construction of the synergistic potential functions on unit quaternion space. Then, we formulate the tracking controller and present the stability analysis.

\subsection{Centrally Synergistic Potential Functions on \texorpdfstring{$\mathbb{S}^3$}{3-Sphere}}
Let $A \in \mathbb{R}^{3\times 3}$ be symmetric and positive definite, and its eigenvalues satisfy $ \lambda_1^A \leq \lambda_2^A \leq \lambda_3^A$ associated with an orthonormal eigenbasis $(v_1^A,v_2^A,v_3^A)$ of $A$. We consider the function \eqref{eq:PF_def} on $\mathbb{S}^3$ with the symmetric matrix $M \coloneqq \mathrm{diag}(0, A) \in \mathbb{R}^{4\times 4}$; that is, 
\begin{equation}
    P(Q) = \epsilon^\top A \epsilon, \label{eq:MTF_quat}
\end{equation} 
where $Q$ is unit quaternion and $\epsilon$ is its vector part. Clearly, $M$ satisfies Assumption \ref{ass:M_cond}. Note that \eqref{eq:MTF_quat} can be written as $P(Q) = \operatorname{tr}(\underline{A} (I - \mathcal{R}_a (Q))) $ with $\underline{A} \coloneqq \frac{1}{4} \operatorname{tr}(A) I - \frac{1}{2} A$, which is so-called \emph{modified trace function} on $\mathrm{SO(3)}$ \cite{Mayhew2013,Berkane2017a,Invernizzi2020}. Since the rigid-body attitude is usually obtained from the body-fixed-frame measurement of the inertial vectors, the parameter $A$ may be determined by using those weighted inertial vectors; see \cite{Berkane2017a}.

Let $\mathbb{Q} = \{1,\dots,6\}$ and define the parameters $\lambda_q$, $u_q$, and $S_q$ for $q \in \mathbb{Q}$ as 
\begin{align}
    \lambda_q &= \lambda_{q+3} = \lambda_q^A &  & q  = \{1,2,3\}, \label{eq:quat_lambda} \\
    u_q &= -u_{q+3} = v_q^A & & q  = \{1,2,3\},
    \label{eq:quat_uq_def} \\
    S_q &= \nu(u_q) \mathbf{i}^\top - \mathbf{i} \nu(u_q)^\top , &  &q  \in \mathbb{Q}.  \label{eq:quat_S_q}
\end{align}
By \eqref{eq:id3} and \eqref{eq:quat_lambda}-\eqref{eq:quat_S_q}, the warping transformation $\mathcal{T}: \mathbb{S}^3 \to \mathbb{S}^3$ of \eqref{eq:warp_def} can be expressed as
\begin{equation}
    \mathcal{T}(Q,q) = 
    \begin{bmatrix}
        \cos (\theta(Q)) & - u_q^\top \sin (\theta(Q)) \\
        u_q \sin (\theta(Q)) & \Pi (u_q) + u_q u_q^\top \cos(\theta(Q))
    \end{bmatrix} Q. \label{eq:warp_quat}
\end{equation}
By Theorem \ref{thm:syn_gap_sol}, we define the warping angle $\theta: \mathbb{S}^3 \to \mathbb{R}_{\geq 0}$ as 
\begin{equation}
    \theta (Q) = k (\lambda_{\max}^A)^{-1} P(Q) = k (\lambda_{\max}^A)^{-1} \epsilon^\top A \epsilon. \label{eq:angle_quat}
\end{equation} 
where $P$ is given by \eqref{eq:MTF_quat} and $0 < k <\frac{\pi}{4}$. Then, the next corollary follows from Theorem \ref{thm:syn_gap_sol}.
\begin{cor} \label{cor:syn_func_quat}
    Consider the function $\bar{\delta}$ given by \eqref{eq:bar_delta_sol} with $M = \mathrm{diag}(0,A) \in \mathbb{R}^{4 \times 4}$ defined as in \eqref{eq:MTF_quat} and $\lambda_q$ given by \eqref{eq:quat_lambda}. Define a function $\delta : \mathbb{Q} \to \mathbb{R}$ such that $0 < \delta (q) < \bar{\delta}(q)$ for all $q \in \mathbb{Q}$. Then, the composition of \eqref{eq:MTF_quat}, \eqref{eq:warp_quat}, and \eqref{eq:angle_quat} given by 
    \begin{equation}
        U(Q,q) = P \bigl( \mathcal{T} (Q,q) \bigr) \label{eq:syn_func_quat}
    \end{equation}
    is centrally synergistic relative to $\{\mathbf{i},-\mathbf{i}\} $ with gap exceeding $\delta$. 
    The gradient of \eqref{eq:syn_func_quat} is given by 
    \begin{align}
        \nabla U (Q,q) &= 2 \biggl( 1 + k \frac{\lambda_q^A}{\lambda_{\max}^A}  \Bigl( \bigl( \eta^2 -  (u_q^\top \epsilon)^2 \bigr) \sin (2\theta(Q) ) \notag  \\
        & \quad + 2\eta (u_q^\top \epsilon) \cos (2 \theta(Q))   \Bigr)  \biggr) \nu (A \epsilon) \notag \\
        & + \lambda_q^A 
        \begin{bmatrix}
            2\eta \sin^2 (\theta(Q)) + u_q^\top \epsilon \sin ( 2\theta(Q)) \\
            \bigl(\eta\sin (2\theta(Q))  - 2u_q^\top \epsilon \sin^2 ( \theta(Q)) \bigr) u_q
        \end{bmatrix} .
        \label{eq:grad_U_quat}
    \end{align}
\end{cor}

\begin{rem}\label{rem:mtf}
    Corollary \ref{cor:syn_func_quat} addresses the challenging problem of generating the centrally synergistic potential functions from a modified trace function on unit quaternions. This problem in the form of $\mathrm{SO(3)}$ has been studied in \cite{Mayhew2013,Berkane2017a,Berkane2017b,Wang2022} with various constraints on the parameter $A$, especially not applicable to the case $\lambda_1^A = \lambda_2^A <\lambda_3^A$. By contrast, our result relaxes the assumption on the parameter $A$, and the parameter $\delta$ can be determined without specifying the inverse trigonometric function about the basic function as the warping angle. Additionally, the proposed angular warping of \eqref{eq:warp_quat} is built on the orthogonal matrix transformation that preserves the structure of $\mathbb{S}^3$, and in consequence, is different from the transformation in \cite{Berkane2017a,Berkane2017b} which takes the form of quaternion multiplication in quaternion group. Furthermore, $U$ of \eqref{eq:syn_func_quat} and its gradient of \eqref{eq:grad_U_quat} are consistent for each rigid-body attitude, since the fact that $U(Q,q) = U(-Q,q)$ and $\nabla U(Q,q) = \nabla U(-Q,q)$ hold for all $(Q,q) \in \mathbb{S}^3 \times \mathbb{Q}$. 
\end{rem}

\subsection{Global Hybrid Attitude Tracking}

Using the synergistic potential functions $U$ and the parameter $\delta$ in Corollary \ref{cor:syn_func_quat}, we define the state feedback $\kappa_2 : \mathbb{S}^3 \times \mathbb{Q} \to \mathbb{R}^3$ by
\begin{align}
    \kappa_2 (\tilde{Q},q) & = \frac{1}{2} \Lambda (\tilde{Q})^\top  \nabla U (\tilde{Q},q) 
\end{align}
where the gradient of $U$ is given by \eqref{eq:grad_U_quat}. It yields the following hybrid controller
\begin{subequations}
    \begin{align}
        & \tau  = \Xi (\tilde{Q},\bar{\omega}_d,\dot{\omega}_d)   -  k_1 \kappa_2 (\tilde{Q},q) - k_2  \tilde{\omega} , \label{eq:ctr_tau} \\
        & 
        \begin{cases}
        \dot{q} = 0 & (\tilde{Q}, q )\in \mathcal{F}_2 ,\\
        q^+ \in G_2 (\tilde{Q},q)  & (\tilde{Q}, q) \in \mathcal{J}_2 , \label{eq:ctr_q}
        \end{cases}
    \end{align} 
    \label{eq:ctr_quat}
\end{subequations}
where $k_1, k_2 > 0$ are constant gains, the jump map is defined as $G_2 (\tilde{Q},q) = \arg \min_{p \in \mathbb{Q}} U(\tilde{Q},p)$, the flow and jump sets are defined as $\mathcal{F}_2 =  \{ (\tilde{Q},q) \in \mathbb{S}^3 \times \mathbb{Q}  : \mu_U (\tilde{Q},q)\leq \delta(q) \} $ and $\mathcal{J}_2 =  \{ (\tilde{Q},q) \in \mathbb{S}^3 \times \mathbb{Q}  : \mu_U (\tilde{Q},q)\geq \delta(q) \} $. 

Define the state space $\mathcal{W}_\xi \coloneqq \mathcal{W}_z \times \mathbb{Q} $ and the state $ \xi \coloneqq (z, q) \in \mathcal{W}_\xi$. The next proposition proved in Appendix \ref{pf:gas_quat} states that the hybrid controller \eqref{eq:ctr_quat} is an effective solution to Problem~\ref{prob:2}.
\begin{prop} \label{prop:gas_quat}
    Problem~\ref{prob:2} is solved by \eqref{eq:ctr_quat} in the sense that the compact set $\mathcal{B}_\xi = \{\xi \in \mathcal{W}_\xi : z \in \mathcal{A}_1 \}$ is globally and robustly asymptotically stable.
\end{prop}
\begin{rem}
    In contrast to \eqref{eq:NonCS_ctr_quat} in \cite{Mayhew2011} and its variants in \cite{Schlanbusch2012,Gui2016,Gui2018,Huang2021,Hashemi2021}, \eqref{eq:ctr_quat} is centrally synergistic and consistent for each rigid-body attitude. It can disambiguate the quaternion measurements and accomplish attitude tracking even when the switching runs into errors, leading to significantly higher robustness from a practical standpoint.
\end{rem}

\section{Simulations} \label{sec:sim}
In this section, numerical examples are presented to illustrate the performance of the proposed centrally synergistic hybrid controller of \eqref{eq:ctr_quat} and we refer to it as `CS Hybrid'. For comparison, we refer to the non-centrally synergistic hybrid controller of \eqref{eq:NonCS_ctr_quat} in \cite{Mayhew2011} as `NonCS Hybrid' and consider the continuous controller, i.e., the $q$-th control law \eqref{eq:ctr_tau} without the hybrid switching mechanism \eqref{eq:ctr_q}, and referred to it as `CS $q = 1$' for $q = 1$. 

\begin{figure}[htb]
    \centering
    \includegraphics[width=5cm]{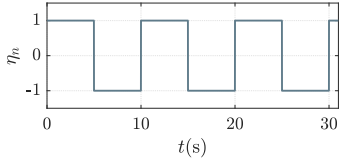}
    \caption{Discontinuous attitude measurement: the scalar part of $Q_n$.} 
    \label{fig:FlipNoise}
\end{figure}

Consider a rigid body with an inertia matrix $J = \mathrm{diag}([0.5,0.7,0.3])\unit{\kilogram.\metre^2}$ to track the reference trajectory generated by \eqref{eq:ref_traj} with initial conditions $Q_d(0) = \mathbf{i}$, $\omega_d (0) = 0 \unit{\radian / \second}$ and $\omega_d(t) = [t e^{-0.5t},0.6\sin(0.4t), 0.6\sin(0.7t)]^\top \unit{\radian / \second} $. The attitude measurement is given by $Q_m = Q \odot Q_n $, where $Q_n = X_{s}(t)[\cos(n_{\alpha}/2),(n_v^\top/ |n_v|)\sin(n_{\alpha}/2) ]^\top$ denotes the attitude noise with zero-mean Gaussian process $n_{\alpha} \sim \mathcal{N}(0,0.01)$ and $n_v \sim \mathcal{N}(0,I_3)$ and the square wave $X_{s}: [0,\infty) \to \{-1,1\}$ that has a period $5\unit{\second}$. Fig.~\ref{fig:FlipNoise} shows that $Q_m$ is periodically discontinuous. The angular-velocity measurement is given by $\omega_m = \omega + n_\omega$ with zero-mean Gaussian process $n_{\omega} \sim \mathcal{N}(0,0.01I_3)$. The gains of the controllers are $k_1 = 4$ and $k_2 = 0.8$. Let $A = \mathrm{diag}([1,1,2])$ in \eqref{eq:MTF_quat} and subsequently $M = \mathrm{diag}([0,A])$. It is noteworthy that the synergistic approaches in \cite{Mayhew2013,Berkane2017a,Berkane2017b,Wang2022} cannot directly apply to this case \eqref{eq:MTF_quat}. For CS Hybrid, $k = 0.5$, $u_q$ is obtained from \eqref{eq:quat_uq_def} with $v_i^A = e_i$ where $e_i$ is the $i$-th column vector of $I_3$, and the synergistic parameter is chosen by $\delta = 0.9 \bar{\delta}$, $\delta(1) = \delta(2) = \delta(4) = \delta(5) = 0.0465$ and $\delta(3) = \delta(6) = 0.0275$.  The synergistic parameter of NonCS Hybrid is set as $\delta = 0.1$.

\begin{figure}[htb]
    \centering
    \includegraphics[width=8cm]{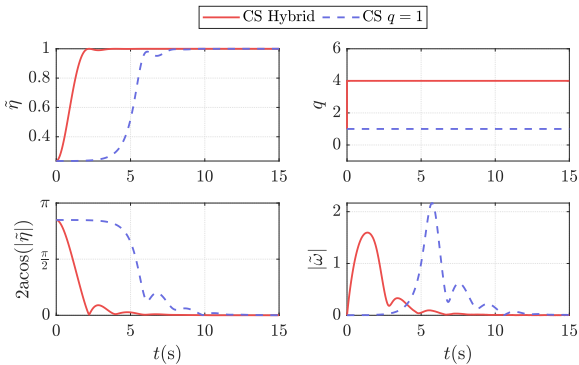}
    \caption{Scenario A: global attitude tracking by CS Hybrid controller.}  
    \label{fig:CSvsCSq1_Global}
\end{figure}

\begin{figure}[htb]
    \centering
    \includegraphics[width=8cm]{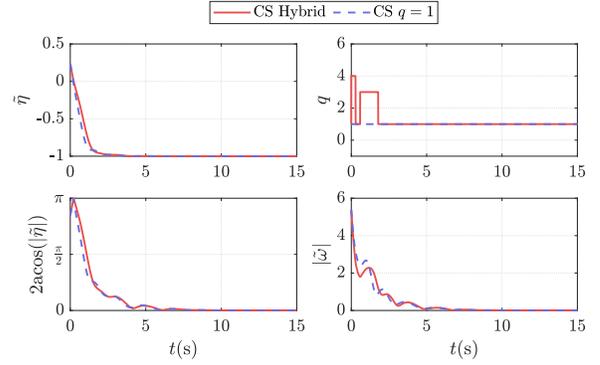}
    \caption{Scenario B: hybrid behavior of CS Hybrid controller.}
    \label{fig:CSvsCSq1_Switch}
\end{figure}

\begin{figure}[htb]
    \centering
    \includegraphics[width=8cm]{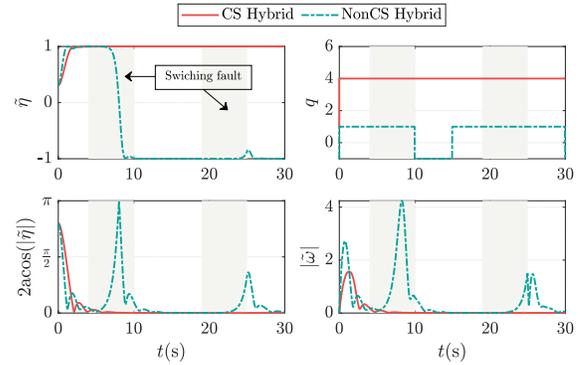}
    \caption{Scenario C: quaternion measurement sensitivity with switching faults (shadow zone).}
    \label{fig:CSvsNonCS}
\end{figure}

The following three scenarios of attitude tracking with different initial conditions are given to illustrate the features of the CS Hybrid.
\begin{enumerate}[A)]
    \item The initial conditions are $Q(0) = [0.2346,0.9721,0,0]^\top$ and $\omega(0) = 0 (\unit{\radian/\second})$. $q(0) = 1$ for CS Hybrid.
    \item The initial conditions are $Q(0) = [0.2346,0.9721,0,0]^\top$ and $\omega(0) = [2,3,4]^\top (\unit{\radian/\second})$. $q(0) = 1$ for CS Hybrid.
    \item The initial conditions are $Q(0) = [0.2346,0.9721,0,0]^\top$ and $\omega(0) = 0 (\unit{\radian/\second})$. $q(0) = 1$ for CS Hybrid and $q(0) = -1$ for NonCS Hybrid. The switching fault occurs during the intervals $(4,10)$ and $(19,25)$, where the switching cannot be triggered.
\end{enumerate}
The simulation results of Scenarios A, B, and C are shown in Figs.~\ref{fig:CSvsCSq1_Global},~\ref{fig:CSvsCSq1_Switch}, and \ref{fig:CSvsNonCS}, respectively. Each figure plots the scalar part $\tilde{\eta}$ of $\tilde{Q}$, the rotation angle $2\mathrm{acos}(|\tilde{\eta}|)$, the norm of the error velocity $\tilde{\omega}$, and the logical variable $q$.

Figure~\ref{fig:CSvsCSq1_Global} shows the \emph{global} tracking feature of the CS Hybrid controller. The initial conditions are set close to the closed-loop undesired equilibrium. Thus, the continuous controller CS $q=1$ exhibits a slow convergence at the onset. This can be overcome by CS Hybrid, because the switching mechanism \eqref{eq:ctr_q} can activate another control law, i.e., $q = 4$ in this case, so as to avoid the feedback vanishing. Similarly, the slow convergence caused by undesired equilibria can also arise in the existing controllers \cite{Akhtar2021,Mason2022,Dong2022}.

In order to show more hybrid behavior, $\tilde{\omega}(0)$ in Scenario B is designed to increase the attitude error under the same $\tilde{Q}(0)$ as Scenario A. As shown in Fig.~\ref{fig:CSvsCSq1_Switch}, since the initial velocity error is too large to decelerate promptly by the controllers, the attitude error moves to zero along a longer path. The CS Hybrid's trajectory moved close enough to the undesired critical points of $U(\tilde{Q},q)$ during the initial period and thus multiple jumps occurred. It is noteworthy that CS Hybrid did not undertake any jump in the end, since the attitude error has already been stabilized close to zero.

Finally, Fig.~\ref{fig:CSvsNonCS} shows that CS Hybrid is impervious to the ambiguity of quaternion measurements and that the tracking goal remains achievable even with switching faults. On the other hand, NonCS Hybrid could overcome the measurement ambiguity depending on the switching mechanism, e.g. the jump of $q$ at $11-19\unit{\second}$. Hence, while the switching mechanism runs into errors, NonCS Hybrid can give an unwinding response to the discontinuous quaternion measurement, e.g., the change of $\tilde{\eta}$ at $4-10 \unit{\second}$. These comparison results demonstrate that CS Hybrid can be more robust than NonCS Hybrid.

\section{Conclusion} \label{sec:con}
In this work, we present a new family of potential functions on $\mathbb{S}^n$ centrally synergistic relative to two antipodal points. It induces a hybrid controller with the explicit expression of the synergistic parameter for global stabilization of a pair of antipodal points on $\mathbb{S}^n$. Furthermore, the result is applicable to global attitude tracking using unit quaternions and is more robust than the existing non-centrally synergistic designs. Additionally, our result relaxes the conservative assumption on the parameter of the basic potential function and can thus handle more control cases.

\section{Acknowledgment}
The authors would like to thank Dr. Pedro Casau for sharing and discussing the results of \cite[Lemma 7 \& 8]{Casau2015a} which motivated Lemma \ref{lem:nec_cond_synFunc}, and would also like to thank the Associate Editor and the anonymous reviewers for their valuable comments.

\appendix

\subsection{Proof of Lemma \ref{lem:nec_cond_synFunc}} \label{pf:nec_cond_synFunc}
By the basic matrix calculus rule, a trivial verification shows that the gradient of \eqref{eq:warp_def} is \eqref{eq:grad_warp}. 

If $\det \bigl( \nabla \mathcal{T}(x,q) \bigr) \neq 0$ for all $(x,q) \in \mathbb{S}^n \times \mathbb{Q}$, the inverse of $\nabla \mathcal{T}(x,q)$ exists and is given by 
\begin{align*}
    \bigl( \nabla \mathcal{T}(x,q) \bigr)^{-1} & = e^{S_q \theta(x)}  \biggl(I + \frac{\nabla \theta (x) x^\top S_q}{1 -  x^\top S_q \nabla \theta (x)} \biggr)
\end{align*} 
where we use the Sherman-Morrison-Woodbury formula \cite[Fact 3.21.3.]{Bernstein2018}. It follows from the inverse function theorem \cite[Thm. 6.26]{Tu2011} that the map $x \mapsto \mathcal{T} (x,q)$ is everywhere a local diffeomorphism.

Since $V$ and $\mathcal{T}$ are differentiable, the composite function $U = V \circ \mathcal{T}$ is differentiable. Noting that $\mathcal{T}^{-1} (\mathcal{A}_0) = \mathcal{A}_0 \times \mathbb{Q} \eqqcolon \mathcal{B}_0 $ by \eqref{eq:setpoint_def} and that $V$ is positive definite relative to $\mathcal{A}_0$, we have that $U$ is positive definite relative to $\mathcal{B}_0$. Using chain rule, we can obtain that for each $(x,q) \in \operatorname{Crit} {U}$, $\Pi (x) \nabla \mathcal{T}(x,q) \nabla V \bigl(\mathcal{T}(x,q)\bigr) = 0$, or equivalently the vector $ \nabla V \bigl(\mathcal{T}(x,q)\bigr) $ either is parallel to $\bigl(\nabla \mathcal{T} (x,q)\bigr)^{-1} x  =  \mathcal{T}(x,q)$ or equals zero since $\nabla \mathcal{T}(x,q) $ has an inverse. It follows that $\Pi (\mathcal{T}(x,q)) \nabla V \bigl(\mathcal{T}(x,q)\bigr) = 0$ for all $ (x,q) \in \operatorname{Crit} { U}$. Therefore, $\operatorname{Crit} {U} = \mathcal{T}^{-1}(\operatorname{Crit} {V})$, as desired.

\subsection{Proof of Theorem \ref{thm:syn_gap_positive}} \label{pf:syn_gap_positive}

By Lemmas \ref{lem:crit_P} and \ref{lem:nec_cond_synFunc}, the set of critical points of $U$ is given by $\operatorname{Crit} {U} = \mathcal{T}^{-1} \bigl( \mathcal{E}_v (M) \bigr)$. It follows from \eqref{eq:warp_def} and \eqref{eq:Sq_def} that for each $(y,q) \in \operatorname{Crit} {U} \setminus \mathcal{B}_0$, there exists $ (\lambda,v ) \in \mathcal{E}(M)$ such that $ \lambda > 0$,
\begin{align} 
    U(y,q) &= \lambda ,  \label{eq:critPoint_def} \\
    y &= 
    \begin{cases}
        v, & \lambda \neq \lambda_q ,\\
        e^{-S_q \theta(y)} v, & \lambda = \lambda_q.
    \end{cases} 
    \label{eq:critPoint_Solut}
\end{align} 

The assumption of $\mathcal{T}^{-1} ( \mathcal{A}_0 ) = \mathcal{B}_0$ guarantees that $y \notin  \mathcal{A}_0 $ and thus $0 < \theta(y) < \pi/4 $. In view of \eqref{eq:critPoint_def}, given $q \in \mathbb{Q}$, we divide the critical point $(y,q)$ of $U$ into two subsets, i.e., $\mathcal{B}_q \coloneqq \{(z,q) \in \operatorname{Crit} {U} : U(z,q) = \lambda_q \}$ and $\mathcal{B}_q^c \coloneqq \{(z,q) \in \operatorname{Crit} {U} \setminus \mathcal{B}_0 : U(z,q) \neq \lambda_q \}$. Now, let us study the synergistic gap in two cases.
\setcounter{case}{0}
\begin{case}
    The first case is $\lambda \notin \{\lambda_q , 0\}$, i.e., $(y,q) \in \mathcal{B}_q^c$. It follows from \eqref{eq:critPoint_Solut} that $y = v$. Since $v$ is an eigenvector of $M$ associated with $\lambda$, we can obtain that for each $p \in \mathbb{Q}_{\lambda}$,
    \begin{align}
        e^{S_p \theta(y) }y & = \Pi (u_p) v + ( v^\top u_p ) \bigl( u_p \cos(\theta(y))  - r \sin (\theta(y)) \bigr) , \notag \\
        U(y,p) &=  \Bigl( 1 - ( v^\top u_p )^2 \sin^2 (\theta(y))\Bigr) \lambda, \label{eq:U_yp_lambda_neq}
    \end{align}
    where Lemma \ref{lem:pre_id}, \eqref{eq:id3}, and $r^\top v = 0$ are used. It follows from \eqref{eq:useful_ineq2}
    \begin{equation}
        U(y,q) - \min_{p \in \mathbb{Q}_\lambda} U(y,p) \geq \frac{\sin^2 (\theta(y))}{\gamma_M (\lambda)}  \lambda .  \label{eq:gap_lambda_notq}
    \end{equation}
    Minimizing both sides of \eqref{eq:gap_lambda_notq} yields that for a given $q \in \mathbb{Q}$
    \begin{align}
        \min_{(y,q) \in  \mathcal{B}_q^c} \mu_U (y,q) & \geq \min_{ \substack{\lambda \in \mathcal{E}_\lambda (M) \\ \lambda \notin \{\lambda_q , 0\} } } \Bigl( U(y,q) - \min_{p \in \mathbb{Q}_\lambda} U(y,p) \Bigr) \geq \Delta_1 (q)  \label{eq:gap_notq}
    \end{align}
    where $0 < \underline{\vartheta} \leq \theta(y) < \frac{\pi}{4}$ has been used and $\Delta_1$ is given by \eqref{eq:bar_delta_def_1}.   
\end{case}
\begin{case}
    The other case is $\lambda = \lambda_q$, i.e., $(y,q) \in \mathcal{B}_q$. We show that there exists the index $p \in \mathbb{Q}_{\lambda_q}$ such that the potential function $U(y,p)$ has a lower value. First, a trivial computation yields that 
    \begin{align}
        e^{-2 S_q \theta(y)} v & = \Pi (u_q) v + ( v^\top u_q ) \bigl( r \sin (2\theta(y))   + u_q \cos (2\theta(y)) \bigr) , \notag \\
        U(y, \mathfrak{q}(q))  &= \Bigl( 1 - ( v^\top u_q )^2 \sin^2 ( 2 \theta (y) )\Bigr) \lambda_q ,\label{eq:U_yp_anti_q}
    \end{align}
    where Lemma \ref{lem:pre_id}, \eqref{eq:id3}, and $r^\top v = u_q^\top r= 0$ have been used with the fact that $u_q = - u_{\mathfrak{q}(q)}$. Next, we consider the two subcases with regard to the geometric multiplicity of $\lambda_q$.
    \begin{subcase}
        If $\gamma_M (\lambda_q) = 1$, it is easily seen that $\mathbb{Q}_{\lambda_q} = \{q, \mathfrak{q}(q)\}$ and $ v \in \{u_q, u_{\mathfrak{q}(q)}\}$ and hence $(v^\top u_q)^2 = 1$. Therefore, combining \eqref{eq:critPoint_def} with \eqref{eq:U_yp_anti_q} yields 
        \begin{equation}
            U(y,q) - \min_{p \in \mathbb{Q}_{\lambda_q}} U(y,p) = \sin^2 (2\theta(y)) \lambda_q .  \label{eq:gap_lambda_q1}
        \end{equation} 
    \end{subcase}
    \begin{subcase}
        If $\gamma_M (\lambda_q) > 1$, then $v$ is any unit vector in the eigenspace of $M$ associated with $\lambda = \lambda_q$. Let $p \in \mathbb{Q}_{\lambda_q} \setminus \{q, \mathfrak{q}(q)\}$. 
        Using $u_p^\top u_q = u_p^\top r = u_q^\top r = 0$, it can be shown that $S_p u_q = 0$, $S_p r = u_p$, and $S_p^2 r = -r$. It follows from \eqref{eq:critPoint_Solut} that
        {\small
        \begin{align}
            & e^{S_p \theta(y)} y = v - u_q (u_q^\top v) \bigl(1 - \cos(\theta(y))\bigr)  \notag \\
            &\qquad - u_p \left(  (u_p^\top v) \bigl(1 - \cos(\theta(y))\bigr) - (u_q^\top v) \sin^2(\theta(y)) \right) \notag \\
            & \qquad + r \left(  (u_q^\top v)  \cos(\theta(y)) - (u_p^\top v) \right) \sin(\theta(y)) , \notag \\
            &U(y,p) = \Bigl( 1 - \Bigl((u_p^\top v) - (u_q^\top v) \cos (\theta(y))  \Bigr)^2 \sin^2 (  \theta (y) )\Bigr) \lambda_q .
        \end{align}}%
        Making use of \eqref{eq:useful_ineq1}, we can obtain that 
        {\small
        \begin{align}
            & U(y,q) - \min_{p \in \mathbb{Q}_{\lambda_q} \setminus \{q, \mathfrak{q}(q)\} } U(y,p) \geq  \notag\\ 
            & \quad  \left( \sqrt{\frac{1 - ( u_q^\top v)^2}{\gamma_M(\lambda_q) - 1}} + \left|u_q^\top v \right| \cos ( \theta (y) )\right)^2 \sin^2 ( \theta (y) ) \lambda_q. \label{eq:tmp1}
        \end{align} } %
        Consider the function $h: [0,1] \to \mathbb{R} $ defined as $h(t) = (a\sqrt{(1-t^2)} + b t)^2$, where $ a \in (0,1)$ and $b \in (0,1]$. We claim that the minimum of $h$ is obtained at either $t =0$ or $t=1$, because $h$ is positive and the function $t \mapsto \sqrt{h(t)}$ is concave. It follows that
        \begin{align}
            \text{LHS of \eqref{eq:tmp1}}  \geq \lambda_q \min \left\{ \frac{\sin^2 ( \theta (y) )}{\gamma_M(\lambda_q) - 1}    ,\frac{\sin^2 ( 2\theta (y))}{4}    \right\}. \label{eq:gap_lambda_q2}
        \end{align}
    \end{subcase}
    Combining \eqref{eq:gap_lambda_q1} and \eqref{eq:gap_lambda_q2} yields
    \begin{align}
    \min_{(y,q) \in \mathcal{B}_q} \mu_U(y,q) \geq U(y,q) - \min_{p \in \mathbb{Q}_{\lambda_q} } U(y,p) \geq \Delta_2 (q), \label{eq:gap_isq}
    \end{align}
    where $\Delta_2$ is given by \eqref{eq:bar_delta_def_2}. 
\end{case}

Note that $\mathcal{B}_q \bigcup \mathcal{B}_q^c$ contain all undesired critical points of $U$ with the index $q$. Therefore, in view of \eqref{eq:gap_notq} and \eqref{eq:gap_isq}, we conclude that
\begin{align*}
    \mu_U (y,q) &\geq \bar{\delta}(q) > 0, & \forall (y,q) &\in \operatorname{Crit} {U} \setminus \mathcal{B}_0
\end{align*}
in which case $\bar{\delta}(q)$ is given by \eqref{eq:bar_delta_def}.
Accordingly, one can always find the function $\delta : \mathbb{Q} \to \mathbb{R}$ satisfying $0 < \delta (q) < \bar{\delta}(q)$, such that $U$ is centrally synergistic relative to $\mathcal{A}_0$ with gap exceeding $\delta$, which completes the proof.

\subsection{Proof of Theorem \ref{thm:syn_gap_sol}} \label{pf:syn_gap_sol}
Using the facts that $\det (I + z_1 z_2^\top) = 1 + z_1^\top z_2$ for all $z_1, z_2 \in \mathbb{R}^n $ \cite[Fact 3.21.3]{Bernstein2018} and that $|z_3^\top z_1 z_2^\top z_3| \leq 0.5$ for all $z_1,z_2 \in \mathbb{S}^n $ with $z_1^\top z_2 = 0$ and all $z_3 \in \mathbb{S}^n$, the determinant of \eqref{eq:grad_warp} satisfies
\begin{align*}
    & \bigl| \det \bigl( \nabla \mathcal{T}(x,q) \bigr) \bigr|  = \bigl| 1 - x^\top S_q \nabla \theta(x) \bigr|    \geq 1 -  k \frac{\lambda_q}{\lambda_n}  > 0 .
\end{align*}
Let $(y,q) \in \mathbb{S}^n \times \mathbb{Q}$ such that $\mathcal{T}(y,q) \in \mathcal{A}_0$. It follows that 
\begin{align}
    P(y) &= P \bigl( e^{-S_q \theta(y)} r \bigr) = \lambda_q \sin^2 \bigl( k \lambda_n^{-1}  P (y)\bigr). \label{eq:crtit_U_zero}
\end{align}
Suppose that $P(y) > 0$. From \eqref{eq:crtit_U_zero}, $0 < k < \frac{\pi}{4} $ and the fact that $\sin^2(z) < z$ for all $z > 0$, we could find that $P(y) < k (\lambda_q / \lambda_n ) P(y) \leq k P(y) < P(y)$, which is impossible. It follows that $P(y) = 0$, hence that $y \in \mathcal{A}_0$ by Assumption \ref{ass:M_cond}, and finally that $\mathcal{T}^{-1} (\mathcal{A}_0  ) = \mathcal{B}_0$. This proves the statement (1).

In view of statement (1) and Lemma \ref{lem:nec_cond_synFunc}, we have that $\operatorname{Crit} {U} = \mathcal{T}^{-1} (\mathcal{E}_v (M))$. Let us evaluate $P$ at the unwanted critical points of $U$. To this end, we use \eqref{eq:critPoint_Solut} to study $P(y)$ by cases as we have done in the proof of Theorem \ref{thm:syn_gap_positive}.
\setcounter{case}{0}
\begin{case}
    If $\lambda \neq \lambda_q$ and $y \in \mathcal{B}_q^c$, then
    \begin{equation}
    P(y) = P(v) = \lambda. \label{eq:Py_notq}
    \end{equation}
\end{case}
\begin{case}
    If $\lambda = \lambda_q$ and $y \in \mathcal{B}_q$, a trivial computation of $P(y) = P(e^{-S_q k \lambda_n^{-1} P(y)} v) $ gives 
    \begin{equation*}
        P(y) = \Bigl( 1 - (v^\top u_q)^2 \sin^2 \bigl(  k \lambda_n^{-1} P(y) \bigr)\Bigr) \lambda_q.
    \end{equation*}
    Making use of $P(y) > 0$, $-1 \leq v^\top u_q \leq 1$, and the fact that $\sin^2 (z) < z^2 $ for all $z > 0$, we can obtain
    \begin{equation}
        \lambda_q \geq P(y) \geq \frac{2  \lambda_q }{1 + \sqrt{1 + 4 (v^\top u_q)^2 k^2 \frac{\lambda_q^2}{\lambda_n^2} }} \geq \Theta (q) \frac{ \lambda_n}{k} . \label{eq:Py_isq}
    \end{equation}
\end{case}

Finally, substituting \eqref{eq:Py_notq} into \eqref{eq:gap_lambda_notq}, and \eqref{eq:Py_isq} into \eqref{eq:gap_lambda_q1} as well as \eqref{eq:gap_lambda_q2}, we can obtain \eqref{eq:bar_delta_sol} in view of the fact that $\mathcal{B}_q \bigcup \mathcal{B}_q^c$ contain all undesired critical points of $U$ for any $q \in \mathbb{Q}$. This proves the statement (2).

\subsection{Proof of Proposition \ref{prop:gas_quat}} \label{pf:gas_quat}
The closed-loop system by \eqref{eq:ref_traj}, \eqref{eq:right_err_traj} and \eqref{eq:ctr_quat} is autonomous and satisfies the hybrid basic conditions \cite[Assumption 6.5]{Goebel2012}. Moreover, its jump map is given by $\xi^+ \in (z , G_2 (\tilde{Q},q))$, and the flow and jump sets are defined as $C_2 = \{ \xi \in \mathcal{W}_\xi:(\tilde{Q}, q )\in \mathcal{F}_2 \}$ and $D_2 = \{\xi \in \mathcal{W}_\xi:(\tilde{Q}, q )\in \mathcal{J}_2 \}$, respectively. Consider the real-valued function $V(\xi) = k_1 U (\tilde{Q},q) + 0.5  \tilde{\omega}^\top J \tilde{\omega}$. Invoking Corollary \ref{cor:syn_func_quat}, the function $U \in \mathcal{C}^1 (\mathbb{S}^3 \times \mathbb{Q}, \mathbb{R})$ of \eqref{eq:syn_func_quat} is centrally synergistic relative to $\{\mathbf{i},-\mathbf{i}\}$. It follows that $V$ is positive definite relative to $\mathcal{B}_\xi$. In view of \eqref{eq:right_err_traj} and \eqref{eq:ctr_quat}, we have that $\dot{V}(\xi) = - k_2 |\tilde{\omega}|^2 \leq 0$ for all $\xi \in C_2$. Additionally, $V(\xi) - V(\xi^+)  = k_1 \mu_U (\tilde{Q},q) > k_1 \delta (q)$ for all $\xi \in D_2 $. Hence, $V$ is nonincreasing along the flows on $C_2 \setminus \mathcal{B}_\xi$ and strictly decreasing over the jumps on $D_2$. Therefore, the set $\mathcal{B}_\xi$ is stable by \cite[Thm. 3.19]{Sanfelice2021}. Additionally, invoking the hybrid invariance principle \cite[Thm. 3.23]{Sanfelice2021}, the solution $\xi$ converges to the largest weakly invariant set in $\dot{V}^{-1}(0) \coloneqq \{ \xi \in C_2 : \dot{V} (\xi) = 0 \}$. Since $\tilde{\omega} = 0$ for $\xi \in \dot{V}^{-1}(0)$, $\kappa_2 \equiv 0$ follows from \eqref{eq:right_err_traj} and \eqref{eq:ctr_tau}, and hence $\Pi (\tilde{Q})\nabla U (\tilde{Q},q) = 0$ follows from the fact that for any $a\in \mathbb{S}^3, b \in \mathbb{R}^4$, $\Lambda(a)^\top b = 0$ if and only if $a$ and $b$ are collinear or $b = 0$. This implies that for $\dot{V}^{-1}(0) \subseteq \mathcal{I} \coloneqq \{z \in C_2 : (\tilde{Q},q) \in \operatorname{Crit} {U}, \tilde{\omega} = 0\}$. Since $U$ is centrally synergistic relative to $\{\mathbf{i},-\mathbf{i}\}$, it follows that $\mathcal{I} \bigcap C_2 = \mathcal{B}_\xi $ and thus that $\mathcal{B}_\xi = \dot{V}^{-1}(0)$. Noting that $\mathcal{B}_\xi$ is invariant, we can conclude that $\mathcal{B}_\xi$ is globally attractive. This proves that $\mathcal{B}_\xi$ is globally asymptotically stable. The robustness of asymptotic stability can be shown similarly in Proposition \ref{prop:gas_sphere}, which completes the proof.

\bibliographystyle{ieeetran}
\bibliography{Reference/references.bib}

\begin{thebibliography}{10}
\providecommand{\url}[1]{#1}
\csname url@samestyle\endcsname
\providecommand{\newblock}{\relax}
\providecommand{\bibinfo}[2]{#2}
\providecommand{\BIBentrySTDinterwordspacing}{\spaceskip=0pt\relax}
\providecommand{\BIBentryALTinterwordstretchfactor}{4}
\providecommand{\BIBentryALTinterwordspacing}{\spaceskip=\fontdimen2\font plus
\BIBentryALTinterwordstretchfactor\fontdimen3\font minus \fontdimen4\font\relax}
\providecommand{\BIBforeignlanguage}[2]{{%
\expandafter\ifx\csname l@#1\endcsname\relax
\typeout{** WARNING: IEEEtran.bst: No hyphenation pattern has been}%
\typeout{** loaded for the language `#1'. Using the pattern for}%
\typeout{** the default language instead.}%
\else
\language=\csname l@#1\endcsname
\fi
#2}}
\providecommand{\BIBdecl}{\relax}
\BIBdecl

\bibitem{Verginis2020}
C.~K. Verginis, M.~Mastellaro, and D.~V. Dimarogonas, ``Robust cooperative manipulation without force/torque measurements: Control design and experiments,'' \emph{{IEEE} Transactions on Control Systems Technology}, vol.~28, no.~3, pp. 713--729, may 2020.

\bibitem{Dong2022}
R.-Q. Dong, A.-G. Wu, and Y.~Zhang, ``Anti-unwinding sliding mode attitude maneuver control for rigid spacecraft,'' \emph{{IEEE} Transactions on Automatic Control}, vol.~67, no.~2, pp. 978--985, feb 2022.

\bibitem{Mason2022}
P.~Mason and L.~Greco, ``Almost global attitude stabilisation of an underactuated axially symmetric 3-d pendulum,'' \emph{Automatica}, vol. 137, p. 110110, mar 2022.

\bibitem{Mattioni2022}
M.~Mattioni, A.~Moreschini, S.~Monaco, and D.~Normand-Cyrot, ``Quaternion-based attitude stabilization via discrete-time {IDA}-{PBC},'' \emph{{IEEE} Control Systems Letters}, vol.~6, pp. 2665--2670, 2022.

\bibitem{Turner2021}
M.~C. Turner and C.~M. Richards, ``Constrained rigid body attitude stabilization: An anti-windup approach,'' \emph{{IEEE} Control Systems Letters}, vol.~5, no.~5, pp. 1663--1668, nov 2021.

\bibitem{Zhang2022}
F.~Zhang, D.~Meng, and X.~Li, ``Robust adaptive learning for attitude control of rigid bodies with initial alignment errors,'' \emph{Automatica}, vol. 137, p. 110024, mar 2022.

\bibitem{Sanfelice2006}
R.~G. Sanfelice, M.~J. Messina, S.~E. Tuna, and A.~R. Teel, ``Robust hybrid controllers for continuous-time systems with applications to obstacle avoidance and regulation to disconnected set of points,'' in \emph{2006 American Control Conference}, 2006, Conference Proceedings, pp. 3352--3357.

\bibitem{Garone2010}
E.~Garone, R.~Naldi, and E.~Frazzoli, ``Switching control laws in the presence of measurement noise,'' \emph{Systems \& Control Letters}, vol.~59, no.~6, pp. 353--364, 2010.

\bibitem{Thienel2003}
{Thienel, J. and Sanner, R. M.}, ``{A coupled nonlinear spacecraft attitude controller and observer with an unknown constant gyro bias and gyro noise},'' \emph{IEEE Transactions on Automatic Control}, vol.~48, no.~11, pp. 2011--2015, 2003.

\bibitem{Mayhew2011}
C.~G. Mayhew, R.~G. Sanfelice, and A.~R. Teel, ``Quaternion-based hybrid control for robust global attitude tracking,'' \emph{IEEE Transactions on Automatic Control}, vol.~56, no.~11, pp. 2555--2566, 2011.

\bibitem{Goebel2012}
R.~Goebel, R.~G. Sanfelice, and A.~R. Teel, \emph{Hybrid Dynamical Systems: Modeling, Stability, and Robustness}.\hskip 1em plus 0.5em minus 0.4em\relax New Jersey: Princeton University Press, 2012.

\bibitem{Sanfelice2021}
R.~Sanfelice, \emph{Hybrid Feedback Control}.\hskip 1em plus 0.5em minus 0.4em\relax New Jersey: Princeton University Press, 2021.

\bibitem{Mayhew2013}
C.~G. Mayhew and A.~R. Teel, ``Synergistic hybrid feedback for global rigid-body attitude tracking on so(3),'' \emph{IEEE Transactions on Automatic Control}, vol.~58, no.~11, pp. 2730--2742, 2013.

\bibitem{Berkane2017a}
S.~Berkane and A.~Tayebi, ``Construction of synergistic potential functions on so(3) with application to velocity-free hybrid attitude stabilization,'' \emph{IEEE Transactions on Automatic Control}, vol.~62, no.~1, pp. 495--501, 2017.

\bibitem{Schlanbusch2012}
R.~Schlanbusch, A.~Loria, and P.~J. Nicklasson, ``On the stability and stabilization of quaternion equilibria of rigid bodies,'' \emph{Automatica}, vol.~48, no.~12, pp. 3135--3141, 2012.

\bibitem{Gui2016}
H.~Gui and G.~Vukovich, ``Global finite-time attitude tracking via quaternion feedback,'' \emph{Systems \& Control Letters}, vol.~97, pp. 176--183, 2016.

\bibitem{Gui2018}
H.~Gui and A.~H.~J. de~Ruiter, ``Global finite-time attitude consensus of leader-following spacecraft systems based on distributed observers,'' \emph{Automatica}, vol.~91, pp. 225--232, 2018.

\bibitem{Huang2021}
Y.~Huang and Z.~Meng, ``Global finite-time distributed attitude synchronization and tracking control of multiple rigid bodies without velocity measurements,'' \emph{Automatica}, vol. 132, p. 109796, 2021.

\bibitem{Hashemi2021}
S.~H. Hashemi, N.~Pariz, and S.~K.~H. Sani, ``Observer-based adaptive hybrid feedback for robust global attitude stabilization of a rigid body,'' \emph{IEEE Transactions on Aerospace and Electronic Systems}, vol.~57, no.~3, pp. 1919--1929, 2021.

\bibitem{Bhat2000}
S.~P. Bhat and D.~S. Bernstein, ``A topological obstruction to continuous global stabilization of rotational motion and the unwinding phenomenon,'' \emph{Systems \& Control Letters}, vol.~39, no.~1, pp. 63--70, 2000.

\bibitem{Mayhew2010}
C.~G. Mayhew and A.~R. Teel, ``Hybrid control of spherical orientation,'' in \emph{49th IEEE Conference on Decision and Control (CDC)}, 2010, Conference Proceedings, pp. 4198--4203.

\bibitem{Mayhew2013b}
------, ``Global stabilization of spherical orientation by synergistic hybrid feedback with application to reduced-attitude tracking for rigid bodies,'' \emph{Automatica}, vol.~49, no.~7, pp. 1945--1957, 2013.

\bibitem{Casau2015a}
P.~Casau, C.~G. Mayhew, R.~G. Sanfelice, and C.~Silvestre, ``Global exponential stabilization on the n-dimensional sphere,'' in \emph{2015 American Control Conference (ACC)}, 2015, Conference Proceedings, pp. 3218--3223.

\bibitem{Casau2019a}
------, ``Robust global exponential stabilization on the n-dimensional sphere with applications to trajectory tracking for quadrotors,'' \emph{Automatica}, vol. 110, p. 108534, 2019.

\bibitem{Raj2021}
N.~Raj, R.~N. Banavar, Abhishek, and M.~Kothari, ``Robust attitude tracking for aerobatic helicopters: A geometric approach,'' \emph{{IEEE} Transactions on Control Systems Technology}, vol.~29, no.~1, pp. 150--164, jan 2021.

\bibitem{Akhtar2021}
A.~Akhtar and S.~L. Waslander, ``Controller class for rigid body tracking on $\mathrm{SO}(3)$,'' \emph{IEEE Transactions on Automatic Control}, vol.~66, no.~5, pp. 2234--2241, 2021.

\bibitem{Invernizzi2020}
D.~Invernizzi, M.~Lovera, and L.~Zaccarian, ``Dynamic attitude planning for trajectory tracking in thrust-vectoring uavs,'' \emph{IEEE Transactions on Automatic Control}, vol.~65, no.~1, pp. 453--460, 2020.

\bibitem{Mayhew2013a}
C.~G. Mayhew, R.~G. Sanfelice, and A.~R. Teel, ``On path-lifting mechanisms and unwinding in quaternion-based attitude control,'' \emph{IEEE Transactions on Automatic Control}, vol.~58, no.~5, pp. 1179--1191, 2013.

\bibitem{Berkane2017b}
S.~Berkane, A.~Abdessameud, and A.~Tayebi, ``Hybrid global exponential stabilization on so(3),'' \emph{Automatica}, vol.~81, pp. 279--285, 2017.

\bibitem{Wang2022}
M.~Wang and A.~Tayebi, ``Hybrid feedback for global tracking on matrix lie groups $so(3)$ and $se(3)$,'' \emph{{IEEE} Transactions on Automatic Control}, vol.~67, no.~6, pp. 2930--2945, jun 2022.

\bibitem{Lee2015}
T.~Lee, ``Global exponential attitude tracking controls on so(3),'' \emph{IEEE Transactions on Automatic Control}, vol.~60, no.~10, pp. 2837--2842, 2015.

\bibitem{Bernstein2018}
D.~S. Bernstein, \emph{Scalar, Vector, and Matrix Mathematics: Theory, Facts, and Formulas}, revised and expanded~ed.\hskip 1em plus 0.5em minus 0.4em\relax Princeton, New Jersey: Princeton University Press, 2018.

\bibitem{Tu2011}
L.~W. Tu, \emph{\BIBforeignlanguage{eng}{An Introduction to Manifolds}}, ser. Universitext.\hskip 1em plus 0.5em minus 0.4em\relax New York, NY: Springer New York, 2011.

\end{thebibliography}

\end{document}